\documentclass[prb,twocolumn,amsmath,showpacs]{revtex4}
\usepackage{graphicx}
\usepackage{amssymb}
\usepackage{citesort}
\usepackage{stmaryrd}

\begin{document}

\title{Quantum critical transition from charge-ordered to superconducting state in the triangular lattice negative-U extended Hubbard model}

\author{S. Mazumdar}
\affiliation{Department of Physics, University of Arizona
Tucson, AZ 85721}
\author{R.T. Clay}
\affiliation{Department of Physics and Astronomy and HPC$^2$ Center for
Computational Sciences, Mississippi State University, Mississippi State MS 39762}
\date{\today}
\begin{abstract}
We demonstrate a robust frustration-driven charge-order to
superconductivity transition in the half-filled negative-U extended
Hubbard model. Superconductivity extends over a broad region of the
parameter space. We argue that the model provides the correct insight
to understanding unconventional superconductivity in the organic
charge-transfer solids and other quarter-filled systems.
\end{abstract}
\maketitle

Spatial broken symmetries such as antiferromagnetism (AFM) and charge
ordering (CO) are proximate to superconductivity (SC) in a number of
exotic systems, including the cuprates \cite{Bednorz86a},
Na$_x$CoO$_2$ $\cdot$ $y$H$_2$O \cite{Takada03b},
$\beta$-Na$_{0.33}$V$_2$O$_5$ \cite{Yamauchi02a} and organic
charge-transfer solids (CTS) such as (TMTCF)$_2$X (here C=S or Se, and
X are closed-shell anions) \cite{Ishiguro}, (BEDT-TTF)$_2$X (hereafter
ET$_2$X) \cite{Ishiguro} and EtMe$_3$Z[Pd(dmit)$_2$]$_2$, Z = P, As
\cite{Kato06}.  Unlike in the cuprates, SC in the CTS is reached not
by doping, but on application of hydrostatic or uniaxial pressure.
CTS crystals often consist of anisotropic triangular lattices of
dimers of the active cationic or anionic molecules.  The average
number of charge carriers $n$ per molecule is $\frac{1}{2}$,
indicating that $n$ per dimer unit cell is 1. Since the $n=1$
triangular lattice provides the classic template for the resonating
valence bond (RVB) electronic structure within the nearest-neighbor
(n.n.) Heisenberg Hamiltonian \cite{Anderson73a}, the idea that spin
frustration drives an AFM-to-SC or spin liquid-to-SC transition in the
CTS has acquired popularity \cite{afmsc}.  Within this picture,
pressure makes the effective anisotropic $n=1$ triangular lattice more
isotropic, and SC occurs over a narrow range of anisotropy between the
more robust AFM and the paramagnetic metal (PM).  Numerical quantum
Monte Carlo \cite{Mizusaki06a} and exact diagonalization \cite{Li08a}
calculations, however, have failed to find superconducting
correlations in the the triangular lattice repulsive $n=1$ Hubbard
model, casting doubt on the mean-field techniques that find SC within
the Hamiltonian.  Experimentally, the situation is complex. (i) SC in
certain CTS is proximate to CO instead of AFM \cite{co-sc}. This has
led to yet other mean-field models with additional Coulomb parameter
of charge-fluctuation mediated SC \cite{Merino01a}. (ii) The
insulating phase proximate to SC in EtMe$_3$[Pd(dmit)$_2$]$_2$ is not
an AFM but a valence bond solid (VBS), with {\it charge
disproportionation} between molecules \cite{Tamura06a}. (iii) AFM is
missing in the insulating state of $\kappa-$ET$_2$Cu$_2$(CN)$_3$ with
a nearly isotropic triangular lattice \cite{Shimizu03a}.  There occur
inhomogeneous charge localization and sharp decrease in spin
susceptibility below 10 K \cite{kawamoto}, which may also be
signatures of a static or fluctuating VBS-like state.  Whether or not
frustration can drive transition to SC from an ordered state therefore
remains an open and intriguing question.

In the present paper we demonstrate a robust frustration-driven SC
within the $n=1$ negative-$U$ extended Hubbard model (EHM) with
n.n. Coulomb repulsion $V$.  Although the literature on the
negative-$U$ Hubbard Hamiltonian is vast, the model has been
investigated primarily for bipartite lattices.  With repulsive $V$,
there is strong tendency to CO in bipartite lattices and SC is absent
\cite{Micnas90}. Frustrated lattices have been investigated within the
Hamiltonian for $V=0$ only \cite{frust}. Now SC dominates and CO is absent.  
Here we begin with the
square lattice with n.n. $V$ and electron hopping, when the ground
state is a checkerboard CO with alternate double occupancies and
vacancies in the square lattice. As the Coulomb interaction and
electron hopping along one diagonal of the square lattice increase
from zero, charge frustration in the emergent triangular lattice leads
%to superconducting pair correlations.  
to first-order transition to a superconducting state. While our primary goal is to
demonstrate the frustration-driven CO-to-SC transition, we also point
out that our work provides insight for understanding unconventional SC
in $n=\frac{1}{2}$ correlated electron systems including the CTS.

We consider the two-dimensional (2D) Hamiltonian,
\begin{multline}
\label{Hamiltonian}
H = - t\sum_{\langle ij \rangle,\sigma}(c_{i,\sigma}^\dagger c_{j,\sigma}+ H.c.) 
-t^{\prime}\sum_{[kl],\sigma}(c_{k,\sigma}^\dagger c_{l,\sigma}+ H.c.)  \\
-|U| \sum_{i} n_{i,\uparrow} n_{i,\downarrow} 
+ V\sum_{\langle ij \rangle} n_in_j + V^{\prime}\sum_{[kl]} n_kn_l
\end{multline}
on an anisotropic triangular lattice. Here $c^{\dagger}_{i,\sigma}$
creates an electron with spin $\sigma$ ($\uparrow$, $\downarrow$) on
site $i$, $n_{i,\sigma}=c^{\dagger}_{i,\sigma}c_{i,\sigma}$, and $n_i
= \sum_{\sigma}n_{i,\sigma}$. $U$ is the on-site Hubbard interaction;
$\langle ... \rangle$ implies n.n. along the x- and y-axes, with
hopping integral and Coulomb repulsion $t$ and $V$,
respectively. Similarly, [$\cdots$] implies neighbors along the
(x+y)-diagonal, with $t^{\prime}$ and $V^{\prime}$ as the hopping and
Coulomb integrals.  We have chosen the same signs for $t$ and
$t^{\prime}$, as the results for $n=1$ are independent of the relative
sign of $t^{\prime}$.  In the following we express all quantities in
units of $t$.  For simplicity we consider mostly $V^{\prime}=V$,
although we have performed calculations also for other $V^{\prime}$.
Nonzero $V^{\prime}$ and $t^{\prime}$ are both crucial for the
CO-to-SC transition.

We have performed exact diagonalizations on a 16-site periodic
lattice.  The quantities we calculate are the structure factor
$S(\vec{Q})$, the bond-order $B_d$ along the $(x+y)$-diagonal, and the
pair correlation function P(r), which are defined as,
\begin{subequations}
\begin{align}
S(\vec{Q})&=\frac{1}{N} \sum_{j,\vec{r}} exp(i\vec{Q} \cdot \vec{r}) \langle (n_j-1)(n_{j+\vec{r}}-1) \rangle \label{eqn-sfac} \\
B_d&= \sum_{\sigma} \langle c^{\dagger}_{j,\sigma} c_{j+\hat{x}+\hat{y}} + H.c. \rangle  \label{Bond-order} \\
P(r)&= \frac{1}{N} \sum_j \langle c^{\dagger}_{j,\uparrow} c^{\dagger}_{j,\downarrow}
c_{j+\vec{r},\downarrow} c_{j+\vec{r},\uparrow}\rangle \label{pair-pair} 
\end{align}
\end{subequations}

We begin with the numerical results for $U=-2$ and $V=V^{\prime}=1$,
for which a first-order like CO-to-SC transition occurs in our finite
lattice at $t^{\prime}_c=0.5$. In Figs.~1(a) and (b) we have plotted
$S(\vec{Q})$ versus $\vec{Q}$ for $t^{\prime}=0.45$ and
$t^{\prime}=0.5$, respectively. The sharp peak at $\vec{Q}=(\pi,\pi)$
in Fig.~1(a) is a signature of the checkerboard CO, with doubly
occupied and vacant sites alternating along the $x$ and $y$-directions
(see Fig.~4(b)). The amplitude of the CO remains practically unchanged
between $t^{\prime}=0$ and 0.48 (see below).  The complete absence of
the $S(\vec{Q})$ peak in Fig.~1(b) indicates a sudden loss of the CO
due to the mobility acquired by the electron pairs. The latter in turn
leads to superconducting pair correlations, as seen in Fig.~1(c),
where we have plotted P(r) against r (where r is in units of the
lattice constant) for both $t^{\prime}=0.45$ and 0.5.  For $r=1.414$,
$P(r)$ has two different values corresponding to diagonals along $x+y$
and $x-y$, respectively. We have chosen to show the smaller of the
two, corresponding to the nonbonded sites along $x-y$, in Fig.~1(c).
The large increase in $P(r)$ for all $r$ (by more than a factor of 2
at n.n., and nearly an order of magnitude at larger r) as $t^{\prime}$
changes from 0.45 to 0.5, as well as the weak dependence of $P(r)$ on
$r$ beyond $r=1$ are signatures of a SC ground state for
$t^{\prime}=0.5$.  While SC is present in the model at electron
densities $n\neq 1$, the CO-to-SC transition is unique to $n=1$.  Our
calculations were done also for $V^{\prime} \neq V$. The ground state
continues to be CO for all $t^{\prime}$ for $V^{\prime}=0$, with the
amplitude nearly independent of $t^{\prime}$ (not shown). Furthermore,
$P(r)$ in this case is nearly 0 for all $r>1$.  Thus both nonzero
$V^{\prime}$ and $t^{\prime}$ are essential for frustration.  We have,
however, determined that the CO-to-SC transition occurs here at nearly
the same $t^{\prime}_c$ for $V^{\prime}=t^{\prime}$.
\begin{figure}
\centerline{\resizebox{1.6in}{!}{\includegraphics{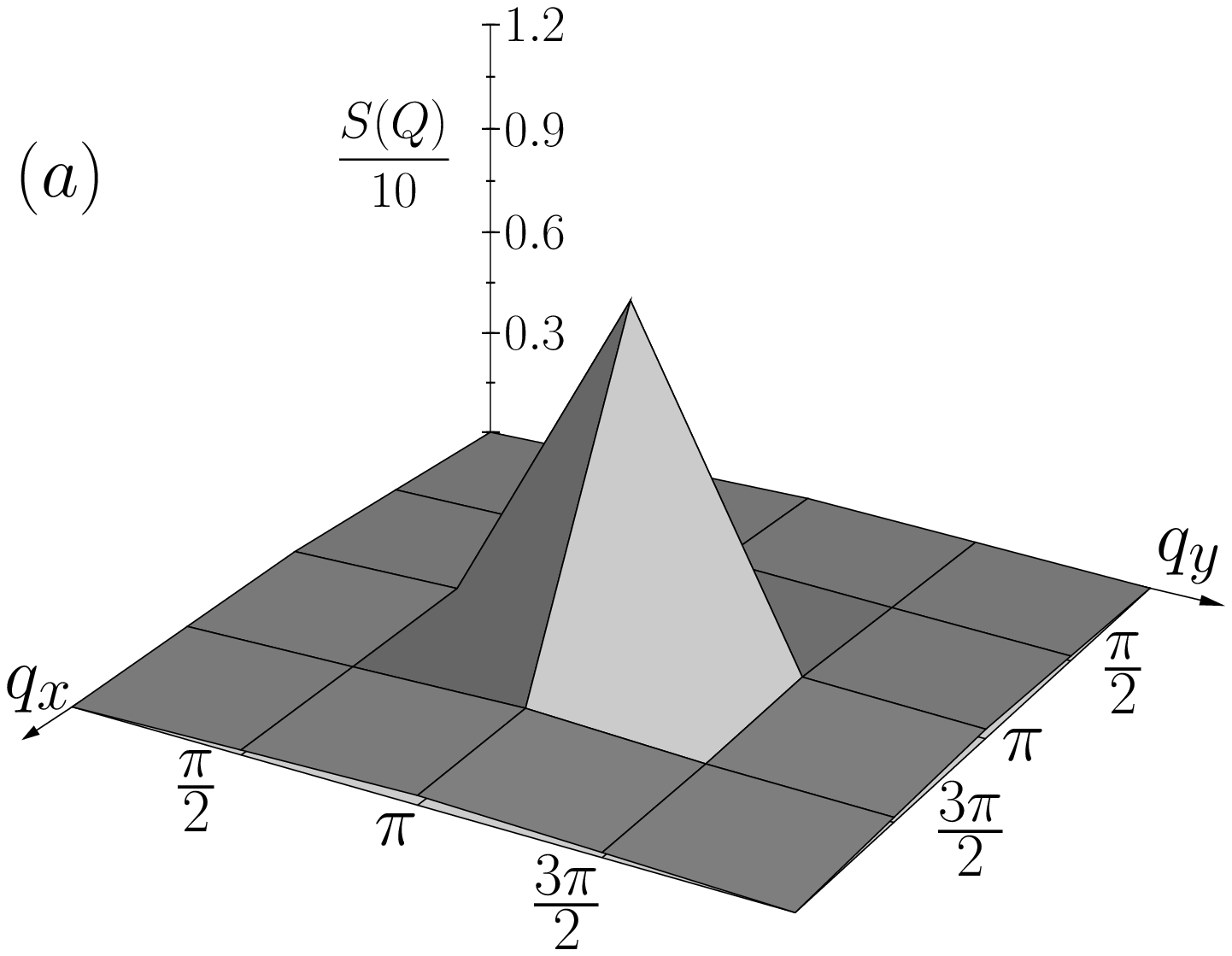}}\resizebox{1.6in}{!}{\includegraphics{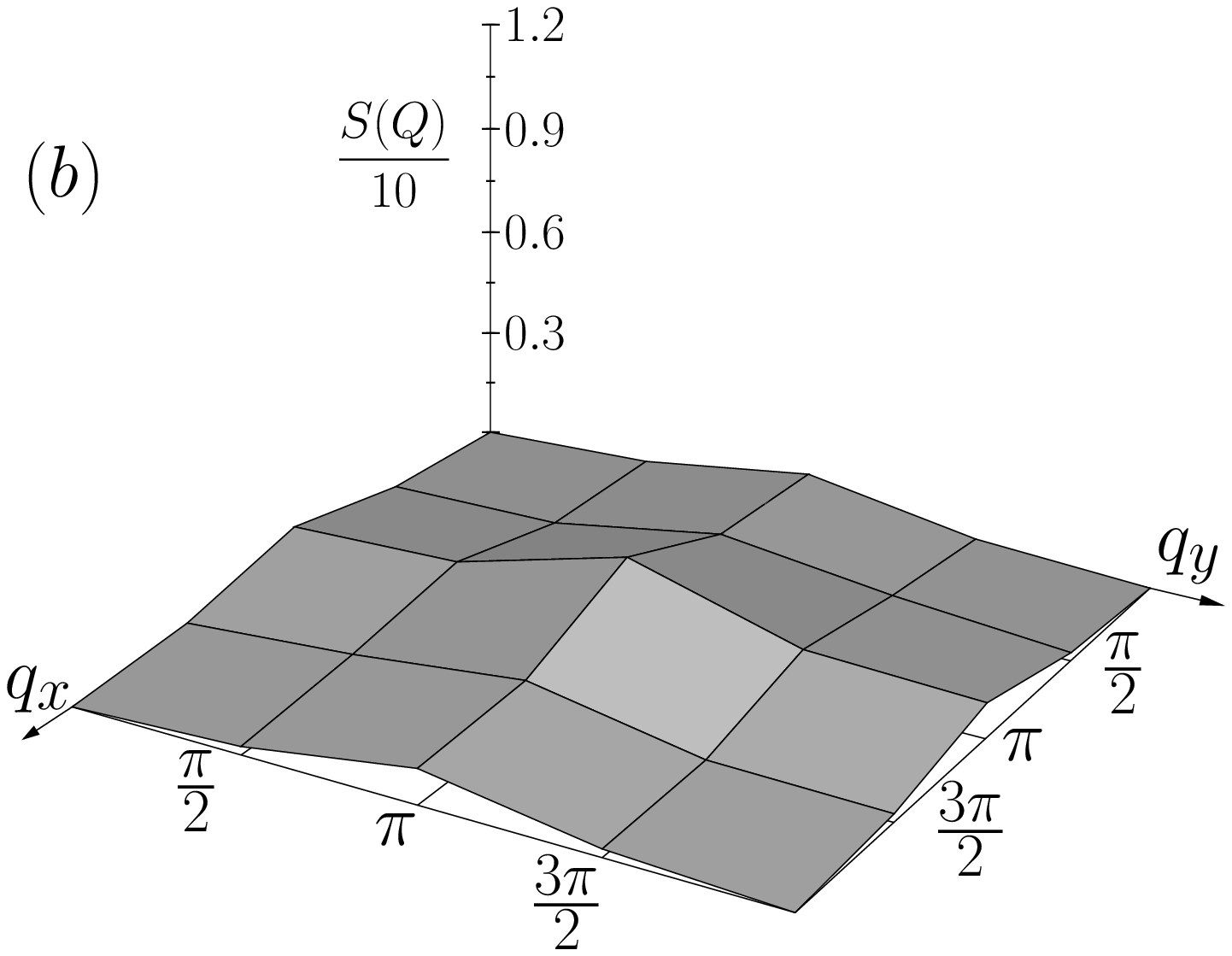}}}
\medskip
\centerline{\resizebox{2.75in}{!}{\includegraphics{fig1c}}}
\caption{(a) and (b), $S(Q)$ vs. $Q$ for $U=-2$, $V=V^{\prime}=1$ and
$t^{\prime}=0.45$ and $t^{\prime}=0.5$, respectively. (c) $P(r)$ vs. $r$,
for $t^{\prime}=0.45$ (dashed curve) and $t^{\prime}=0.5$ (solid curve).}
\end{figure}

To illustrate the sharpness of the CO-to-SC transition in this case we
have plotted in Fig.~2(a) $S(\pi,\pi)$ and $B_d$ against $t^{\prime}$
for the same $U, V$ as in Fig.~1.  The sudden drop in $S(\pi,\pi)$ and
the jump in $B_d$ occur at the same $t^{\prime}_c$, indicating that
the vanishing of the CO is due to sudden increase in carrier mobility.
Fig.~2(b) shows a plot of $P(r_{max}=2.428)$, where the jump in the
pair-correlation occurs at the same $t^{\prime}_c$, indicating that
charge mobility is due to pair motion.  The results of Fig.~2 indicate
that coexistence of CO and SC occurs over a very narrow region of the
parameter space, if at all, for this moderately strongly correlated
case.  This is in contrast to the mixed CO-SC state that is obtained
when instead of $t^{\prime}$ and $V^{\prime}$ the carrier
concentration $n$ is varied \cite{Micnas90}.
\begin{figure}
 \centering
 \includegraphics[clip,width=2.75in]{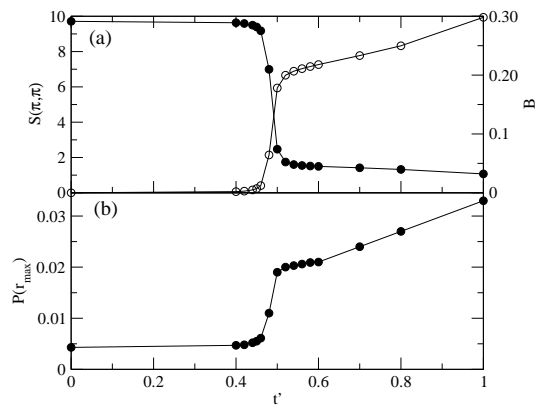}
 \caption{(a) $S(\pi,\pi)$ (filled circles) and $B_d$ (unfilled circles) vs. 
$t^{\prime}$ for $U=-3$,
$V=V^{\prime}=1$. (b) $P(r_{max})$ vs. $t^{\prime}$ for the same 
Coulomb interactions.
}
\end{figure}

Within Eq.~1, the amplitude of the CO at $t^{\prime}=0$ increases with
both $|U|$ and $V$.  As might be expected, the stronger the CO in the
$t^{\prime}=0$ limit, the larger is the $t^{\prime}_c$ at which the
CO-to-SC transition occurs.  Furthermore, it is known that for $V=0$
the CO and SC states are degenerate in the square lattice.  It is then
to be expected that the transition is second order for weak $V$,
independent of the value of $|U|$.  We show a $t^{\prime}-|U|$ phase
diagram for fixed $V=1$ in Fig.~3(a). The transition remains first
order like for this moderate $V$ until $|U|$ is large.  Our phase
diagram in the $t^{\prime}-V$ space in Fig.~3(b) is for fixed
$|U|=4$. Surprisingly, when both $|U|$ and $V$ are large ($|U|=4$,
$V=2$) the ground state continues to be the checkerboard CO for
$t^{\prime}$ as large as 0.9.
\begin{figure}
 \centering
 \includegraphics[clip,width=2.75in]{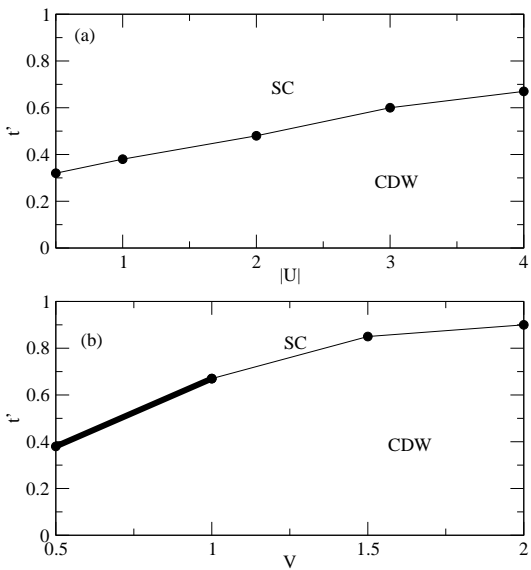}
 \caption{Phase diagrams of the frustrated EHM at zero temperature in
(a) the $t^{\prime}-|U|$ plane for fixed $V=1$; (b) the $t^{\prime}-V$
plane for fixed $U=-4$. The transition is first order (continuous) for
large (small) V in (b). The SC phase persists in the $t^{\prime}>1$
region until $t^{\prime}\gg 1$.  }
\end{figure}

The SC phase within Eq.~\ref{Hamiltonian} is robust and occurs over a
broad region in the phase diagrams in Figs.~3(a) and (b). For the
$|U|$ values of interest SC is lost only for $t^{\prime} \gg 1$. This
is in sharp contrast to the results obtained within the spin
frustration model with repulsive $U$, where even within the
approximate methods that claim SC, the bulk of the $t^{\prime}-U$
parameter space is occupied by AFM and PM phases, and SC occurs as an
intermediate phase for a narrow range of $t^{\prime}$ for each $U$
\cite{afmsc}.

We now speculate how the negative-U EHM in the weak $|U|$ limit may be
relevant to the CTS, $\beta$-Na$_{0.33}$V$_2$O$_5$ and other
$n=\frac{1}{2}$ systems.  Our fundamental premise is that the behavior
of EtMe$_3$Z[Pd(dmit)$_2$]$_2$ and $\kappa-$ET$_2$Cu$_2$(CN)$_3$ with
nearly isotropic lattices is representative of the more anisotropic
systems at very low temperatures and under pressure, when $V^{\prime}$
and $t^{\prime}$ are large.  The mapping of the molecular
$n=\frac{1}{2}$ lattice into the effective $n=1$ repulsive Hubbard
Hamiltonian that describes the AFM phase requires {\it homogeneous}
charge population on the dimer unit cells \cite{afmsc}.  We postulate
that under pressure the AFM phase switches over to a static or
fluctuating VBS with {\it inhomogenous} charge distribution (see
below), and the effective $n=1$ Hamiltonian that describes this state
is different.  We rationalize this hypothesis based on our recent
work.  The complete Hamiltonian for these systems must start from the
$n=\frac{1}{2}$ {\it repulsive}-$U$ EHM.  In a series of papers
\cite{bcdw}, we have established that the ground state of the
$n=\frac{1}{2}$ repulsive Hamiltonian is often a Bond-Charge-Density
Wave (BCDW), with charge occupancy $\cdots1100\cdots$, where `1' and
`0' refer to molecular charges $0.5+\epsilon$ and $0.5-\epsilon$,
respectively. This is a {\it quantum} effect driven by the
antiferromagnetic correlations due to the repulsive $U$ and dominates
over the classical effect due to $V$ that favors the formation of the
Wigner crystal $\cdots1010\cdots$ for $V<V_c(U)$, where $V_c(U)$ can
be as large as 3 for realistic $U$ \cite{bcdw}.  The BCDW is enhanced
by the intersite electron-phonon interactions that modulate the
hopping integral, and the intrasite Holstein electron-molecular
vibration couplings \cite{bcdw}. Most importantly, even when at high
temperatures there occurs a Wigner crystal CO in the (TMTTF)$_2$X, the
low temperature spin-Peierls phase is the BCDW \cite{bcdw}.  We
postulate that the $\cdots1100\cdots$ is a {\it bipolaron density
wave} that can be modeled by an {\it effective} $n=1$ negative-U EHM,
where the effective sites are alternating pairs of occupied (1-1) and
unoccupied (0-0) molecules, respectively.  That the sites of a
negative-U EHM for a complex system can be composite ones has been
suggested previously \cite{Micnas90}. Note that the traditional
mapping of $n=\frac{1}{2}$ into effective $n=1$ AFM \cite{afmsc} also
assumes composite dimer sites, with the only difference that our
proposed mapping is valid for the inhomogenous charge distribution.
\begin{figure}
 \centering
 \includegraphics[clip,width=2.65in]{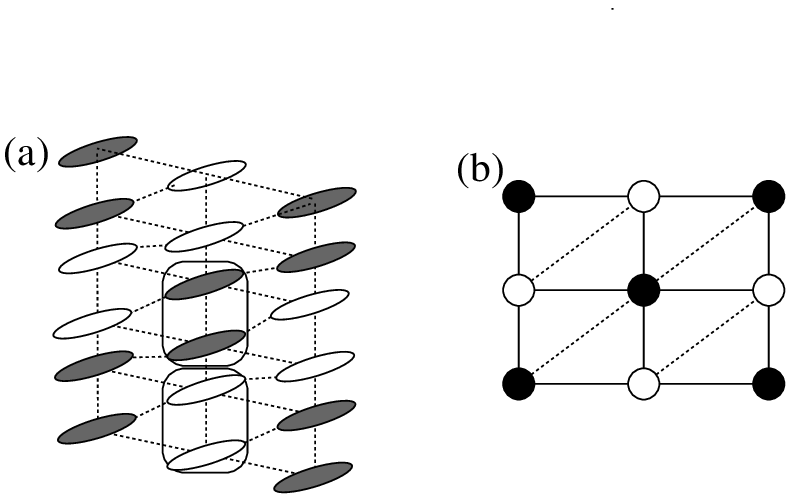}
 \caption{(a) BCDW corresponding to the VBS in the 2D organic
layer. Filled (unfilled) ellipses correspond to charge-rich
(charge-poor) molecules. (b) Pairs of charge-rich and charge-poor
molecules constitute the double occupancy (filled circle) and vacancy
(empty circle) in the effective $n=1$ lattice.  }
\end{figure}

The $\cdots1100\cdots$ charge occupancy persists for dimensionality
greater than 1 \cite{bcdw2d}.  Experimentally, the 2D BCDW has been
seen in the weakly 2D (TMTSF)$_2$X, the triangular lattice
$\theta$-ET$_2$X, the ladder materials (DTTTF)$_2$M(mnt)$_2$ (M=Au,
Cu), and EtMe$_3$[Pd(dmit)$_2$]$_2$.  The co-existing 2k$_F$ charge-
and spin-density waves \cite{Pouget97a} in (TMTSF)$_2$X has been
explained as a Bond-Charge-Spin Density Wave \cite{Mazumdar99a}.  The
experimentally observed horizontal charge stripe in the
$\alpha$-ET$_2$X and $\theta$-ET$_2$X consists of $\cdots1100\cdots$
charge occupancies occurring along the two directions of strong
hoppings \cite{bcdw2d}.  The large spin gap in the two-chain
(DTTTF)$_2$M(mnt)$_2$ \cite{Ribas05a} can be understood within the
$n=\frac{1}{2}$ zigzag ladder model \cite{zigzag}, within which there
occur interchain singlets and the $\cdots1100\cdots$ charge occupancy
along the zigzag diagonals \cite{zigzag}.  We point out that charge
occupancies and the bonding patterns in the VBS
EtMe$_3$P[Pd(dmit)$_2$]$_2$ (see Fig.~3(b) in
Ref. \onlinecite{Tamura06a} and Fig.~4) {\it correspond precisely to
the $\cdots1100\cdots$ BCDW.}  Finally, $\beta$-Na$_{0.33}$V$_2$O$_5$
consists of $n=\frac{1}{2}$ V chains and ladders, and there is
extensive literature on n.n. intersite bipolarons in this and related
systems \cite{Chakraverty78a} (the earlier literature did not
emphasize the $n=\frac{1}{2}$ carrier concentration that is crucial
for the mapping into the effective $n=1$ Hamiltonian).

We therefore postulate that $n=\frac{1}{2}$ systems at low
temperatures and under pressure can be understood qualitatively within
the effective $n=1$ negative-$U$ EHM. This ``mapping'' is seen in
Fig.~4, where we have shown the intersite bipolaron CDW corresponding
to the VBS in EtMe$_3$P[Pd(dmit)$_2$]$_2$ \cite{Tamura06a}, as well as
the corresponding $n=1$ checkerboard CO. We speculate that for small
$t^{\prime}$ within the $n=\frac{1}{2}$ molecular system, the
occupancies of the dimer unit cells are indeed homogeneous, which
gives the AFM in this region. The larger $t^{\prime}$ or lattice
distortion (or both) at low temperatures give the insulating VBS with
CO, and at still larger $t^{\prime}$ the intersite spin singlets
become mobile, which is approximately modeled in our calculations
within the effective negative-$U$ EHM. We emphasize that intersite (as
opposed to intrasite) bipolarons have been shown to be light and
mobile, {\it especially in the triangular lattice} \cite{bipolaron},
and will be even more so here as the binding is due to
antiferromagnetic correlations and not overscreening of $V$ by
electron-phonon interactions. Finally, although SC within the
effective Hamiltonian is $s$-wave, this need not be true within the
actual $n=\frac{1}{2}$ repulsive-$U$ EHM with intersite pairs.

In summary, we have shown that it is indeed possible to have
frustration-driven SC, within a model that emphasizes charge as
opposed to spin-correlations. We have also suggested that the
$n=\frac{1}{2}$ VBS insulator can be conceived as an effective $n=1$
bipolaron CDW. A complete theory of SC in the CTS should of course
demonstrate the CO-to-SC transition within the actual $n=\frac{1}{2}$
repulsive-$U$ EHM with electron-phonon interactions, but we believe
that our work on the effective $n=1$ Hamiltonian gives the insight
necessary for constructing such a theory. In particular, the approach
opens up the possibility of describing all CTS as well as inorganic
$n=\frac{1}{2}$ systems within one simple model. An interesting aspect
of our work is that the configuration space pairing in the VBS, and by
implication in the superconducting state within the present model
occurs due to a co-operative interaction between AFM correlations and
electron-phonon interactions, and thus SC mediated by these two
interactions need not be mutually exclusive.

This work was supported by the Department of Energy grant DE-FG02-06ER46315.

\end{document}